\documentclass[sigconf]{acmart}

\AtBeginDocument{%
  }

\setcopyright{acmlicensed}
\copyrightyear{2024} 
\acmYear{2024} 
\acmConference[RecTemp @ RecSys '24]{Eighteenth ACM Conference on Recommender Systems}{October 14--18, 2024}{Bari, Italy}
\acmBooktitle{18th ACM Conference on Recommender Systems (RecSys '24), October 14--18, 2024, Bari, Italy}

\acmDOI{}
\acmPrice{}
\acmISBN{}
\renewcommand\footnotetextcopyrightpermission[1]{%
  \footnotetext{
\vspace{0.5em} % Adds vertical space before the rule
\hrule % Horizontal line
\noindent\\
Permission to make digital or hard copies of all or part of this work for personal or
classroom use is granted without fee provided that copies are not made or distributed
for profit or commercial advantage and that copies bear this notice and the full citation
on the first page. Copyrights for third-party components of this work must be honored.
For all other uses, contact the owner/author(s). \\
\textit{Presented at Temporal Reasoning in Recommender workshop (RecTemp) @ RecSys ’24, October 14–18, 2024, Bari, Italy} \\
© 2024 Copyright held by the owner/author(s). \\
Correspondence to: Dwipam Katariya <dwipam.katariya@capitalone.com>
}
}

\begin{document}

\title{TIMeSynC: Temporal Intent Modelling with Synchronized Context Encodings for Financial Service Applications}

\author{Dwipam Katariya}
\authornote{Equal Contribution}
\orcid{0009-0009-1058-1244}
\affiliation{%
  \institution{Capital One}
  \city{McLean}
  \state{VA}
  \country{USA}
}
\author{Juan Manuel Origgi}
\authornotemark[1]
\orcid{0000-0003-3617-6019}
\affiliation{%
  \institution{Capital One}
  \city{New York}
  \state{NY}
  \country{USA}
}
\author{Yage Wang}
\authornotemark[1]
\orcid{0009-0008-5738-0987}
\affiliation{%
  \institution{Capital One}
  \city{McLean}
  \state{VA}
  \country{USA}
}
\author{Thomas Caputo}
\orcid{0009-0000-4247-9740}
\affiliation{%
  \institution{Capital One}
  \city{McLean}
  \state{VA}
  \country{USA}
}

\renewcommand{\shortauthors}{Katariya et al.}

\begin{abstract}
  Users engage with financial services companies through multiple channels, often interacting with mobile applications, web platforms, call centers, and physical locations to service their accounts. The resulting interactions are recorded at heterogeneous temporal resolutions across these domains. This multi-channel data can be combined and encoded to create a comprehensive representation of the customer's journey for accurate intent prediction. This demands sequential learning solutions. NMT transformers achieve state-of-the-art sequential representation learning by encoding context and decoding for the next best action to represent long-range dependencies. However, three major challenges exist while combining multi-domain sequences within an encoder-decoder transformers architecture for intent prediction applications: a) aligning sequences with different sampling rates b) learning temporal dynamics across multi-variate, multi-domain sequences c) combining dynamic and static sequences. We propose an encoder-decoder transformer model to address these challenges for contextual and sequential intent prediction in financial servicing applications. Our experiments show significant improvement over the existing tabular method.
\end{abstract}

\begin{CCSXML}
<ccs2012>
   <concept>
       <concept_id>10010147.10010257</concept_id>
       <concept_desc>Computing methodologies~Machine learning</concept_desc>
       <concept_significance>500</concept_significance>
       </concept>
   <concept>
       <concept_id>10002951.10003317.10003347.10003350</concept_id>
       <concept_desc>Information systems~Recommender systems</concept_desc>
       <concept_significance>500</concept_significance>
       </concept>
   <concept>
       <concept_id>10010147.10010178.10010187</concept_id>
       <concept_desc>Computing methodologies~Knowledge representation and reasoning</concept_desc>
       <concept_significance>300</concept_significance>
       </concept>
   <concept>
       <concept_id>10002951.10002952.10002953.10010820.10010518</concept_id>
       <concept_desc>Information systems~Temporal data</concept_desc>
       <concept_significance>500</concept_significance>
       </concept>
 </ccs2012>
\end{CCSXML}

\ccsdesc[500]{Computing methodologies~Machine learning}
\ccsdesc[500]{Information systems~Recommender systems}
\ccsdesc[300]{Computing methodologies~Knowledge representation and reasoning}
\ccsdesc[500]{Information systems~Temporal data}

\keywords{sequential recommendation, tabular synthesis, transformers, temporal learning, heterogeneous data}

\maketitle
\section{Introduction}
Companies in the financial services industry have adopted a multi-channel approach to enable customers to service their accounts including shopping at stores. User behavior across all channels can be used to deliver a personalized digital experience at the moment of sign-on as well as in-session on digital platforms for targeted marketing, increased find-ability, contextual chatbot Q\&A, and more. 
In financial services, the data is stored in tabular format \cite{creditriskscor}, but it also exhibits strong sequential properties. Unlike tabular data where rows are independent, sequential tabular data exhibits dependent rows with dependent columns \cite{fata-trans, unititab}. This data exhibits strong dynamic and static patterns capturing local transient patterns and global identity \cite{tabbert}. For example, the customer's click-stream activity and the amount spent exhibit a dynamic pattern, while the type of product owned exhibits a static pattern replicated across the sequence for an elongated period. Another property of multi-domain data is to be stored at the granular level of the respective timeline as illustrated in Figure \ref{fig:customer journey}. These properties make an important representation of a customer's state for their next best action. It’s a common practice to perform feature engineering and convert raw sequences into the tabular data \cite{trxnclass, deeptlf} for machine learning algorithms such as Gradient Boosted Trees \cite{trxnclass, deeptlf, creditcardfraud}. However, extensive feature engineering is required on sequential tabular data leading to long hours spent in developing the accurate features. Similarly, methods for representing the customer's next action based on the customer's sequence of historical interactions are well explored \cite{SASRec, bert4rec, pinnerformer, rnntopk, genrec, trans4rec, sessionrec, cntxt-awr, seqgnn}. For contextual sequential recommendation, transformers have often shown superior performance over other architectures \cite{SASRec, bert4rec, trans4rec, gtransrec}. However, due to the cross-correlation between sequences and miss-alignment, out-of-the-box transformer architecture cannot be directly applied. Another challenge is to incorporate time information into the sequence to better understand customer’s seasonal, periodic, and temporal behavior across the domains.

\section{Relevant Work}

\subsection{Sequential Recommendation}

The evolution of sequential recommendation systems has been driven by advancements in deep learning, particularly in the ability to model user behavior over time. Sequential recommendation models initially focused on capturing patterns in user behavior using techniques like Markov Chains (MC) and Recurrent Neural Networks (RNN) but struggles with long-term dependencies and computational efficiency. SASRec \cite{SASRec} introduced self-attentive attention to capturing both short-term actions and long-term user preferences, lacking a bi-directional representation of the events. BERT4Rec \cite{bert4rec} extended SASRec by introducing masked loss and allowing the model to represent future actions. Transformer-based models showed success in some industries where recommendation system applications have high business value, most evident in the e-commerce sector, where user behavior data is abundant and continually evolving; for instance Alibaba's Behavior Sequence Transformer (BST) \cite{alibab}, Pinterest's TransAct \cite{transact, pinnerformer}, LinkedIn's LiGNN \cite{lignn} and many others. These innovations address practical challenges like scalability, latency, and cost, but do not do enough to address heterogeneous data within the contextual recommendation framework where heterogeneous data is abundant.

\subsection{Handling Tabular Data with LLM}

Recent advancements in large language models (LLMs) have opened new opportunities for handling structured data, notably in tabular data synthesis and understanding. TaBert \cite{tabert} extends the capabilities of BERT to tabular data, by introducing a content snapshot mechanism, that efficiently encodes relevant table segments to align with natural language queries, significantly improving processing speed and accuracy. TabuLa \cite{tabula} introduces a novel synthesis framework for tabular data, by employing a specialized compression strategy and a unique padding method and addressing the efficiency and quality of synthetic data generation. Finally, the Table-GPT \cite{tablgpt} framework represents a significant leap in the table-specific tuning of generative pre-trained transformers. Together, these models illustrate the transformative impact of LLMs in the field of tabular data handling. Even though they allow for more efficient table representation, to our knowledge there are no benchmarks available for representing heterogeneous data in financial services applications for sequential recommendation.
\begin{figure}
    \centering
    \includegraphics[scale=0.33]{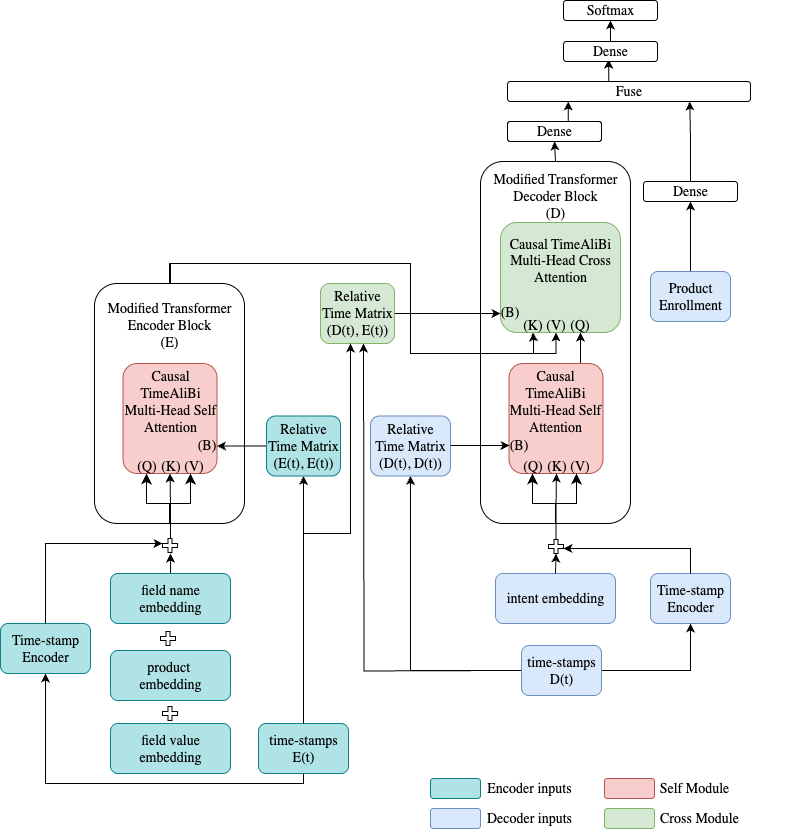}
    \caption{Model Architecture. Field Value embeddings are added with Product and Field Name embeddings before employing the encoder block. For both the decoder and encoder, a Time encoder and TimeAliBi are applied for learning absolute and relative time dynamics. Depending upon the self-attention or cross-attention, a causal mask(K(t), Q(t)) is applied with respect to the time corresponding to Key(K) and the Query(Q). In blue the encodings used as input of the Encoder block; in Red the TimeAliBi self-attention module; in purple the input encodings to the Decoder module; in yellow and green the decoder TimeAlibi cross-attention modules, and at last in white the head layers. This figure only notes changes w.r.t. to the vanilla transformers \cite{vaswani}}
    \label{fig:model}
\end{figure}

\subsection{Sequential Tabular Learning and Representation}
The importance of tabular data sequences as a common data storage type to represent financial transactions recorded in a bank database \cite{unititab}, has led researchers to improve modeling \cite{tabnet}. TabNET \cite{tabnet} introduced a deep learning architecture focused on tabular structures. It incorporates sequential attention outperforming both traditional machine learning and early deep learning. TabTransformer \cite{ft_transformer} expands TabNET by integrating a Transformer Encoder block for domain-specific field embeddings, followed by FT-Transformer to contextualize continuous and numerical features. However, it did not sufficiently address the sequential aspect of the data. TabBERT \cite{tabbert}, adapted from the BERT architecture \cite{bert}, is a hierarchical architecture that utilizes self-attention mechanisms to capture complex relationships between features and temporal dependencies in sequential tabular data, demonstrating improved performance on financial forecasting and customer behavior prediction tasks. FATA-Trans \cite{fata-trans} further extends this approach by isolating static tabular features from dynamic sequential data, enabling better differentiation between static and dynamic fields, reducing negative effects from static field replication, and improving temporal information learning. Similarly, UnitTAB \cite{unititab} extends TabBERT by incorporating row-type dependent embedding to handle variable data dimensions. The model \cite{tabert, unititab, fata-trans} are reported to be used for multiple classification tasks such as pollution prediction, fraud detection, and loan default prediction. 

In this context, we propose a novel approach called TIMeSynC (Temporal Intent Modelling with Synchronized Context Encodings) for contextual recommendations in financial services that incorporates the following items: 
    \begin{itemize}
        \item First flatten and tokenize the customer's multi-channel activity as an encoder context and incorporate a sequence of online intent as a decoder for the next intent prediction. 
        \item ALiBi \cite{alibi} based time representation and time encoder for efficient learning of the temporal cross-correlation between multivariate and multi-domain sequences. 
        \item A time-based attention mask for cross-attention in the decoder module is employed to dynamically attend causal tokens for learning temporal patterns between intent and the context.
        \item Field Name embeddings are added to the encoder for domain and field-aware representation. 
        \item Static sequences of financial product enrollment are fused for point-in-time representation with the decoder output. Product embeddings are added to the encoder sequence for product-aware representation.
    \end{itemize}
    These additional components eliminate the need for multi-level hierarchical modeling, simplifying the architecture to learn complex, deeper, and richer patterns by direct interaction between cross-field and cross-domain dependencies over time.
    
    \begin{table}[htb]
        \caption{Sample intents}
        \label{tab:intent space}
        \resizebox{\columnwidth}{!}{
          \begin{tabular}{c c c}
            \toprule
            Make a payment & Report Fraud & Cancel Zelle Payment \\
            Redeem rewards & Add External Account & ... \\
            Enroll into paperless & Apply Contactless Card & ... \\
            Inquire about Benefits & Cancel Balance Transfer & Update Birthday \\
            \bottomrule
          \end{tabular}
        }
    \end{table}

    \begin{table}[htb]
        \caption{Sample data streams used for this study}
        \label{tab:domain space}
        \resizebox{\columnwidth}{!}{
          \begin{tabular}{c c}
            \toprule
            Domains & Description \\
            \midrule
             Transactions & Credit Card transactions \\
             Payments & Credit Card payments \\
             Rewards & Rewards redeemed and earned activities \\
             Outbound messages & Servicing messages to the customer \\
             ... & ... \\
             Product Enrollment & Point in time customer product enrollment \\
            \bottomrule
          \end{tabular}
        }
    \end{table}

\section{Problem Statement}
The sequential intent prediction problem in finance aims to forecast user actions based on inputs from diverse domains. Let \( X = [X_1, X_2, \ldots, X_n] \) represent \( n \) domains, where \( X_i \) denotes different types of financial interactions, such as credit card/debit card transactions, payments, rewards, account status, customer care messages, call center intents, digital page views, etc. Each \( X_i \) is a time series represented as \( [X_{i1}, X_{i2}, \ldots, X_{it}] \). The goal is to predict future user's intents \( Y = [y_1, y_2, \ldots, y_k] \) at time \( t \), given the financial context \( X \) and the prior sequence of user's intents \( Y \).

\begin{figure}
    \centering
    \includegraphics[scale=0.5]{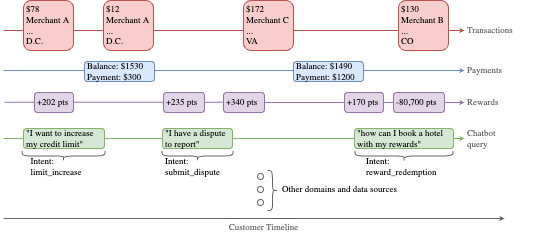}
    \caption{Illustrative customer journey across domains at their respective timelines}
    \label{fig:customer journey}
\end{figure}

\begin{figure}
    \centering
    \includegraphics[scale=0.26]{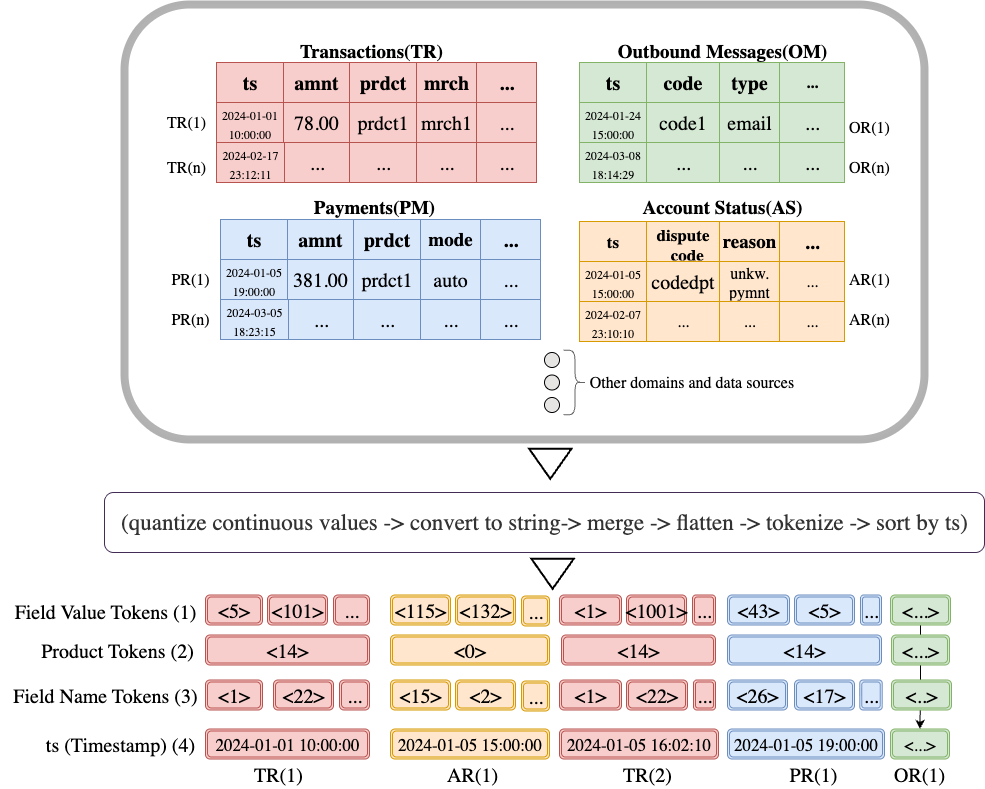}
    \caption{Illustrative Example. Data Flattening and Tokenization of the Encoder Context}
    \label{fig:data process}
\end{figure}

\section{Experiments}

\subsection{Experimental Setup}
    We use various data sources from the internal data ecosystem as mentioned in Table \ref{tab:domain space}. Each data source is associated with its respective timestamps. Therefore, we can construct a flat sequence of customer’s activities all arranged by their respective timestamp. An example of a customer timeline is shown in Figure \ref{fig:customer journey}. We then temporally split the intent and context sequence per user into training/validation/test for model training. We focus our research on 30M+ user sequences with at least one financial activity in the past, and using 500+ intents as supervision. Table \ref{tab:intent space} shows the sample intent space for this study.

    Finally, Assuming we have a set of user U for each intent I, we compute for evaluation the top k recall as:
    \[
        \text{Recall@k}(u) = \frac{\left| \text{Intents}(u) \cap \text{Top}_k(\text{Ranked\_Intents}(u)) \right|}{\left| \text{Intents}(u) \right|}
    \]

\subsection{Model Framework}
    \subsubsection{Baseline(s)} 
    \begin{table*}
      \caption{Offline relative lift in ranking recall over SASRec expressed in \%. Ranking recall is calculated by sorting softmax scores in descending order}
      \label{tab:model recalls}
      \begin{tabular}{c c c c}
        \toprule
        Methods & Recall@1 & Recall@5 & Recall@10\\
        \midrule
        \texttt{SASRec}& 0.00 & 0.00 & 0.00 \\
        \texttt{SASRec + Tabular Context}& \underline{+2.60} & \underline{+2.23} & \underline{+0.03} \\
        \texttt{SASRec + Encoder(Tabular Context)}& +4.23, \underline{+1.59} & +4.06, \underline{+1.79} & +0.98, \underline{+0.95} \\
        \texttt{TIMeSynC}& \textbf{+7.81}, \underline{+3.43} & \textbf{+5.54}, \underline{+1.43} & \textbf{+2.18}, \underline{+1.19} \\
        \bottomrule
      \end{tabular}
    \end{table*}
    
        \begin{itemize}
            \item SASRec \cite{SASRec}: a self-attention based causal recommender, that estimates the importance of the sequence itself without the context. We employ learned positional embeddings as described in \cite{SASRec}. 
            \item SASRec + Tabular Context: This is our current method. We perform extensive feature engineering across different data domains to represent customer's financial context state over time in tabular format. These features are aligned with the timestamp of the intent. We then fuse these features with the SASRec output before the final loss.
            \item SASRec + Encoder(Tabular Context): To measure existing gaps in our feature engineering, we add an encoder block for the tabular features. We then adapt causal encoder-decoder architecture from \cite{vaswani}. For the encoder module, a causal time mask is applied based on the context timestamp. For the cross-attention, a causal time mask is applied based on context and intent timestamp.
        \end{itemize}

\begin{table}[htb]
  \caption{Relative negative effect of each feature when excluded from training expressed in \%}
  \label{tab:model recalls}
  \resizebox{\columnwidth}{!}{
  \begin{tabular}{c c c c c}
    \toprule
    & Feature & Recall@1\\
    \midrule
    \texttt{Full Model}&  & 100.00 \\
    \texttt{Decoder}& w/o Time-stamp Encoder & -0.227 \\
    \texttt{Decoder}& w/o TimeAliBi Self Attn. & -0.361 \\
    \texttt{Decoder}& w/o TimeAliBi Cross Attn. & -0.715 \\
    \texttt{Decoder}& w/o Product Embedding & -1.246 \\
    \texttt{Encoder}& w/o Field Name Embedding & -1.272 \\
    \texttt{Encoder}& w/o Time-stamp Encoder & -0.815  \\
    \texttt{Encoder}& w/o TimeAliBi Self Attn. & -1.082  \\
    \texttt{Encoder}& w/o Product Embeddings & -1.338 \\
    \bottomrule
  \end{tabular}
  }
\end{table}

\subsubsection{TIMeSynC:} 
    For each data source, we first quantile-discretize each numerical field in bins and convert it to string. Then we combine and collapse different data sources together. Finally, we sort them by their respective timestamps as summarized in Figure \ref{fig:data process}. This results in X(u) with u the User and n the sequence length: 
    \[[(t_i,... ,t_n), (fn_i,... ,fn_n), (fv_i,... ,fv_n), (p_i,... ,p_n)]\]
    Where t is the timestamp, fn is the field name, fv is the field value, and p is the product. All fields in X are then string-tokenized except the timestamp. Similarly, sequences of intents are sorted by the timestamp are then string tokenized resulting in Y(u): 
    \[[(t_i,... ,t_n), (y_i,... ,y_n)]\]
    Where t is the timestamp, y is the intent. The model architecture is visualized in Figure \ref{fig:model} and described below:
   \begin{itemize} 
    \item \textbf{TimeAliBi($t(m)$, $t(n)$)}: Inspired by AliBi \cite{alibi}, we modify the attention mechanism \cite{vaswani} to include a bias based on relative time. Before computing the final attention scores, a relative time matrix is added to the attention scores, followed by applying the softmax function for score normalization. Formally, we define:
    \[
    \text{softmax} \left( \frac{ q_i K^{\mathrm{T}} + s \cdot [ t(q_i) - t(k_j) \mid i \in \{1, \ldots, n\}, j \in \{1, \ldots, m\} ] }{\sqrt{d_k}} \right) \cdot V
    \]
    where $V$ is the value matrix, $q_i$ is the $i$-th query, $K$ is the key matrix, $n$ is the query sequence length, $m$ is the key sequence length, and $s$ is the slope parameter.

    \item \textbf{Context Encoder}: The Transformer Encoder Layer, as described in \cite{vaswani}, processes a raw sequence of field values ($S$). A temporal causal mask is then applied based on relative time. To capture sequence periodicity, seasonality, and frequent behavior, time representations based on multi-dimensional signals such as the day of the week, week of the month, and hour of the day are added to the sequence embeddings as described in \cite{amazontime}. The TimeAliBi($E(t)$, $E(t)$) function is then used to capture the relativity of time, where $E(t)$ denotes the time associated with the encoder sequence.

    \item \textbf{Intent Decoder}: The Transformer Decoder Layer, as described in \cite{vaswani}, processes a raw sequence of intents. We modify the decoder layer as follows: The cross-attention mechanism uses Queries ($q_i$) from the intent self-attention, Keys ($k_j$), and Values ($v_j$) from the output of the context encoder ($E$). TimeAliBi ($D(t)$, $E(t)$) is applied to represent the relativity of time, followed by a causal mask $\text{causal\_mask} (D(t), E(t))$, where $D(t)$ denotes the time associated with the decoder sequence. Similar to the encoder, time representations are also incorporated in the decoder as shown in Figure \ref{fig:model}.

    \item \textbf{Field Name Embeddings}: Due to flattening and tokenization, domain and field representations can be lost for certain quantities (e.g., amount). To retain this information, we encode field and domain names as \texttt{<field\_name\_+\_domain>}. These are then string-tokenized and embedded before being added to the raw sequence.

    \item \textbf{Product Representation}: Users can own multiple accounts, reflecting their interactions with those accounts. To incorporate product awareness in the encoder, product embeddings are added to the raw sequence of the encoder. Additionally, to identify point-in-time product ownership in the decoder module, the product one-hot encoding is first passed through a dense layer and then fused with the final decoder output.
\end{itemize}

\subsection{Results}
Table \ref{tab:model recalls} shows a performance comparison between TIMeSynC and baselines. We specifically consider short-term ranking recall@1,5,10 due to the business needs and the UI specifications. The tabulated values are expressed as \%, the best results are boldfaced and the relative lift from the prior row is underlined. SASRec provides an important baseline for understanding the value of a sequence of intents and establishing a lower bound. Both point-in-time and sequential context with the encoder demonstrate superior performance over SASRec showing the importance of tabular context. TIMeSynC outperforms SASRec and further shows improved performance over SASRec + Encoder(Tabular Context) proving the loss of signal in prior methods.
\textit{Feature Ablation:} In Table \ref{tab:model recalls}, we see the impact of each feature on the final model performance by removing it from training. The two features that contribute noticeably are the Product embedding and Field Name embedding. The temporal aspects are of higher importance in the encoder context than the decoder block. Given the importance of each feature, we choose to include all features in TIMeSynC.

\section{Conclusion and Future Work}
In this paper, we proposed a novel framework for context-aware sequential recommendation in financial services applications for effectively combining heterogeneous data with sequential actions. Specifically, we tackle this problem by leveraging an encoder-decoder architecture and flattening across the domain, field, and time. We also employ TimeAliBi and a multi-dimensional time encoder for representing absolute and relative time. Our empirical results demonstrate the gap in the existing feature-engineered tabular context and highlight the significance of our approach. \textit{Limitation(s):} We acknowledge that our approach for encoding context has some constraints compared to a hierarchical framework approach, due to the flattening across the domain, field, and time that could lead to an explosion of the encoder context window. Additionally, we acknowledge that the results may not generalize to other datasets that do not exhibit the same characteristics. \textit{In the Future:} We aim to improve numerical representation, and tokenization and incorporate other data sources such as click-stream, call center, and credit bureau data. We further aim to apply TIMeSynC to other recommendation objectives (vehicle and marketing recsys, mobile app, shopping, and business deals personalization). \textit{Broader Impact:} While our approach is primarily studied in the context of user's Q\&A intent prediction, it may apply to other prominent objectives in financial services such as fraud, call center intent, and charge-off prediction. Our work could apply to other industries like e-commerce, entertainment, and tourism given the peculiarities and differences in data.

\begin{acks}
We would like to thank Arturo Hernandez Zeledon who contributed to the discussions and reviewed the paper. We would like to thank Qi Yu and Giri Iyengar who supported us through this study.
\end{acks}

\bibliographystyle{ACM-Reference-Format}
\bibliography{sample-base}


\begin{thebibliography}{29}

%%% the Web address.

\ifx \showCODEN    \undefined \def \showCODEN     #1{\unskip}     \fi
\ifx \showDOI      \undefined \def \showDOI       #1{#1}\fi
\ifx \showISBNx    \undefined \def \showISBNx     #1{\unskip}     \fi
\ifx \showISBNxiii \undefined \def \showISBNxiii  #1{\unskip}     \fi
\ifx \showISSN     \undefined \def \showISSN      #1{\unskip}     \fi
\ifx \showLCCN     \undefined \def \showLCCN      #1{\unskip}     \fi
\ifx \shownote     \undefined \def \shownote      #1{#1}          \fi
\ifx \showarticletitle \undefined \def \showarticletitle #1{#1}   \fi
\ifx \showURL      \undefined \def \showURL       {\relax}        \fi
% invisible to TeX
\providecommand\bibfield[2]{#2}
\providecommand\bibinfo[2]{#2}
\providecommand\natexlab[1]{#1}
\providecommand\showeprint[2][]{arXiv:#2}

\bibitem[Arik and Pfister(2021)]%
        {tabnet}
\bibfield{author}{\bibinfo{person}{Sercan~{\"O} Arik} {and} \bibinfo{person}{Tomas Pfister}.} \bibinfo{year}{2021}\natexlab{}.
\newblock \showarticletitle{Tabnet: Attentive interpretable tabular learning}. In \bibinfo{booktitle}{\emph{Proceedings of the AAAI conference on artificial intelligence}}, Vol.~\bibinfo{volume}{35}. \bibinfo{pages}{6679--6687}.
\newblock


\bibitem[Borisov et~al\mbox{.}(2023)]%
        {deeptlf}
\bibfield{author}{\bibinfo{person}{Vadim Borisov}, \bibinfo{person}{Klaus Broelemann}, \bibinfo{person}{Enkelejda Kasneci}, {and} \bibinfo{person}{Gjergji Kasneci}.} \bibinfo{year}{2023}\natexlab{}.
\newblock \showarticletitle{DeepTLF: robust deep neural networks for heterogeneous tabular data}.
\newblock \bibinfo{journal}{\emph{International Journal of Data Science and Analytics}} \bibinfo{volume}{16}, \bibinfo{number}{1} (\bibinfo{year}{2023}), \bibinfo{pages}{85--100}.
\newblock


\bibitem[Borisyuk et~al\mbox{.}(2024)]%
        {lignn}
\bibfield{author}{\bibinfo{person}{Fedor Borisyuk}, \bibinfo{person}{Shihai He}, \bibinfo{person}{Yunbo Ouyang}, \bibinfo{person}{Morteza Ramezani}, \bibinfo{person}{Peng Du}, \bibinfo{person}{Xiaochen Hou}, \bibinfo{person}{Chengming Jiang}, \bibinfo{person}{Nitin Pasumarthy}, \bibinfo{person}{Priya Bannur}, \bibinfo{person}{Birjodh Tiwana}, {et~al\mbox{.}}} \bibinfo{year}{2024}\natexlab{}.
\newblock \showarticletitle{Lignn: Graph neural networks at linkedin}. In \bibinfo{booktitle}{\emph{Proceedings of the 30th ACM SIGKDD Conference on Knowledge Discovery and Data Mining}}. \bibinfo{pages}{4793--4803}.
\newblock


\bibitem[Cao and Lio(2024)]%
        {genrec}
\bibfield{author}{\bibinfo{person}{Panfeng Cao} {and} \bibinfo{person}{Pietro Lio}.} \bibinfo{year}{2024}\natexlab{}.
\newblock \showarticletitle{GenRec: Generative Personalized Sequential Recommendation}.
\newblock \bibinfo{journal}{\emph{arXiv preprint arXiv:2407.21191}} (\bibinfo{year}{2024}).
\newblock


\bibitem[Chang et~al\mbox{.}(2021)]%
        {seqgnn}
\bibfield{author}{\bibinfo{person}{Jianxin Chang}, \bibinfo{person}{Chen Gao}, \bibinfo{person}{Yu Zheng}, \bibinfo{person}{Yiqun Hui}, \bibinfo{person}{Yanan Niu}, \bibinfo{person}{Yang Song}, \bibinfo{person}{Depeng Jin}, {and} \bibinfo{person}{Yong Li}.} \bibinfo{year}{2021}\natexlab{}.
\newblock \showarticletitle{Sequential recommendation with graph neural networks}. In \bibinfo{booktitle}{\emph{Proceedings of the 44th international ACM SIGIR conference on research and development in information retrieval}}. \bibinfo{pages}{378--387}.
\newblock


\bibitem[Chen et~al\mbox{.}(2019)]%
        {alibab}
\bibfield{author}{\bibinfo{person}{Qiwei Chen}, \bibinfo{person}{Huan Zhao}, \bibinfo{person}{Wei Li}, \bibinfo{person}{Pipei Huang}, {and} \bibinfo{person}{Wenwu Ou}.} \bibinfo{year}{2019}\natexlab{}.
\newblock \showarticletitle{Behavior sequence transformer for e-commerce recommendation in alibaba}. In \bibinfo{booktitle}{\emph{Proceedings of the 1st international workshop on deep learning practice for high-dimensional sparse data}}. \bibinfo{pages}{1--4}.
\newblock


\bibitem[Chen et~al\mbox{.}(2024)]%
        {gtransrec}
\bibfield{author}{\bibinfo{person}{Yi-Cheng Chen}, \bibinfo{person}{Yen-Liang Chen}, {and} \bibinfo{person}{Chia-Hsiang Hsu}.} \bibinfo{year}{2024}\natexlab{}.
\newblock \showarticletitle{G-TransRec: A Transformer-Based Next-Item Recommendation With Time Prediction}.
\newblock \bibinfo{journal}{\emph{IEEE Transactions on Computational Social Systems}} (\bibinfo{year}{2024}).
\newblock


\bibitem[de~Souza Pereira~Moreira et~al\mbox{.}(2021)]%
        {trans4rec}
\bibfield{author}{\bibinfo{person}{Gabriel de Souza Pereira~Moreira}, \bibinfo{person}{Sara Rabhi}, \bibinfo{person}{Jeong~Min Lee}, \bibinfo{person}{Ronay Ak}, {and} \bibinfo{person}{Even Oldridge}.} \bibinfo{year}{2021}\natexlab{}.
\newblock \showarticletitle{Transformers4rec: Bridging the gap between nlp and sequential/session-based recommendation}. In \bibinfo{booktitle}{\emph{Proceedings of the 15th ACM conference on recommender systems}}. \bibinfo{pages}{143--153}.
\newblock


\bibitem[Devlin(2018)]%
        {bert}
\bibfield{author}{\bibinfo{person}{Jacob Devlin}.} \bibinfo{year}{2018}\natexlab{}.
\newblock \showarticletitle{Bert: Pre-training of deep bidirectional transformers for language understanding}.
\newblock \bibinfo{journal}{\emph{arXiv preprint arXiv:1810.04805}} (\bibinfo{year}{2018}).
\newblock


\bibitem[Gorishniy et~al\mbox{.}(2021)]%
        {ft_transformer}
\bibfield{author}{\bibinfo{person}{Yury Gorishniy}, \bibinfo{person}{Ivan Rubachev}, \bibinfo{person}{Valentin Khrulkov}, {and} \bibinfo{person}{Artem Babenko}.} \bibinfo{year}{2021}\natexlab{}.
\newblock \showarticletitle{Revisiting deep learning models for tabular data}.
\newblock \bibinfo{journal}{\emph{Advances in Neural Information Processing Systems}}  \bibinfo{volume}{34} (\bibinfo{year}{2021}), \bibinfo{pages}{18932--18943}.
\newblock


\bibitem[Hidasi and Karatzoglou(2018)]%
        {rnntopk}
\bibfield{author}{\bibinfo{person}{Bal{\'a}zs Hidasi} {and} \bibinfo{person}{Alexandros Karatzoglou}.} \bibinfo{year}{2018}\natexlab{}.
\newblock \showarticletitle{Recurrent neural networks with top-k gains for session-based recommendations}. In \bibinfo{booktitle}{\emph{Proceedings of the 27th ACM international conference on information and knowledge management}}. \bibinfo{pages}{843--852}.
\newblock


\bibitem[Kang and McAuley(2018)]%
        {SASRec}
\bibfield{author}{\bibinfo{person}{Wang-Cheng Kang} {and} \bibinfo{person}{Julian McAuley}.} \bibinfo{year}{2018}\natexlab{}.
\newblock \showarticletitle{Self-Attentive Sequential Recommendation}. In \bibinfo{booktitle}{\emph{2018 IEEE International Conference on Data Mining (ICDM)}}.
\newblock


\bibitem[Kotios et~al\mbox{.}(2022)]%
        {trxnclass}
\bibfield{author}{\bibinfo{person}{Dimitrios Kotios}, \bibinfo{person}{Georgios Makridis}, \bibinfo{person}{Georgios Fatouros}, {and} \bibinfo{person}{Dimosthenis Kyriazis}.} \bibinfo{year}{2022}\natexlab{}.
\newblock \showarticletitle{Deep learning enhancing banking services: a hybrid transaction classification and cash flow prediction approach}.
\newblock \bibinfo{journal}{\emph{Journal of big Data}} \bibinfo{volume}{9}, \bibinfo{number}{1} (\bibinfo{year}{2022}), \bibinfo{pages}{100}.
\newblock


\bibitem[Luetto et~al\mbox{.}(2023)]%
        {unititab}
\bibfield{author}{\bibinfo{person}{Simone Luetto}, \bibinfo{person}{Fabrizio Garuti}, \bibinfo{person}{Enver Sangineto}, \bibinfo{person}{Lorenzo Forni}, {and} \bibinfo{person}{Rita Cucchiara}.} \bibinfo{year}{2023}\natexlab{}.
\newblock \showarticletitle{One transformer for all time series: Representing and training with time-dependent heterogeneous tabular data}.
\newblock \bibinfo{journal}{\emph{arXiv preprint arXiv:2302.06375}} (\bibinfo{year}{2023}).
\newblock


\bibitem[Moreira et~al\mbox{.}(2021)]%
        {sessionrec}
\bibfield{author}{\bibinfo{person}{Gabriel de Souza~P Moreira}, \bibinfo{person}{Sara Rabhi}, \bibinfo{person}{Ronay Ak}, \bibinfo{person}{Md~Yasin Kabir}, {and} \bibinfo{person}{Even Oldridge}.} \bibinfo{year}{2021}\natexlab{}.
\newblock \showarticletitle{Transformers with multi-modal features and post-fusion context for e-commerce session-based recommendation}.
\newblock \bibinfo{journal}{\emph{arXiv preprint arXiv:2107.05124}} (\bibinfo{year}{2021}).
\newblock


\bibitem[Padhi et~al\mbox{.}(2021)]%
        {tabbert}
\bibfield{author}{\bibinfo{person}{Inkit Padhi}, \bibinfo{person}{Yair Schiff}, \bibinfo{person}{Igor Melnyk}, \bibinfo{person}{Mattia Rigotti}, \bibinfo{person}{Youssef Mroueh}, \bibinfo{person}{Pierre Dognin}, \bibinfo{person}{Jerret Ross}, \bibinfo{person}{Ravi Nair}, {and} \bibinfo{person}{Erik Altman}.} \bibinfo{year}{2021}\natexlab{}.
\newblock \showarticletitle{Tabular transformers for modeling multivariate time series}. In \bibinfo{booktitle}{\emph{ICASSP 2021-2021 IEEE International Conference on Acoustics, Speech and Signal Processing (ICASSP)}}. IEEE, \bibinfo{pages}{3565--3569}.
\newblock


\bibitem[Pancha et~al\mbox{.}(2022)]%
        {pinnerformer}
\bibfield{author}{\bibinfo{person}{Nikil Pancha}, \bibinfo{person}{Andrew Zhai}, \bibinfo{person}{Jure Leskovec}, {and} \bibinfo{person}{Charles Rosenberg}.} \bibinfo{year}{2022}\natexlab{}.
\newblock \showarticletitle{Pinnerformer: Sequence modeling for user representation at pinterest}. In \bibinfo{booktitle}{\emph{Proceedings of the 28th ACM SIGKDD conference on knowledge discovery and data mining}}. \bibinfo{pages}{3702--3712}.
\newblock


\bibitem[Press et~al\mbox{.}(2021)]%
        {alibi}
\bibfield{author}{\bibinfo{person}{Ofir Press}, \bibinfo{person}{Noah~A Smith}, {and} \bibinfo{person}{Mike Lewis}.} \bibinfo{year}{2021}\natexlab{}.
\newblock \showarticletitle{Train short, test long: Attention with linear biases enables input length extrapolation}.
\newblock \bibinfo{journal}{\emph{arXiv preprint arXiv:2108.12409}} (\bibinfo{year}{2021}).
\newblock


\bibitem[Qiu et~al\mbox{.}(2019)]%
        {creditriskscor}
\bibfield{author}{\bibinfo{person}{Ziyue Qiu}, \bibinfo{person}{Yuming Li}, \bibinfo{person}{Pin Ni}, {and} \bibinfo{person}{Gangmin Li}.} \bibinfo{year}{2019}\natexlab{}.
\newblock \showarticletitle{Credit risk scoring analysis based on machine learning models}. In \bibinfo{booktitle}{\emph{2019 6th International Conference on Information Science and Control Engineering (ICISCE)}}. IEEE, \bibinfo{pages}{220--224}.
\newblock


\bibitem[Rahmani et~al\mbox{.}(2023)]%
        {amazontime}
\bibfield{author}{\bibinfo{person}{Mostafa Rahmani}, \bibinfo{person}{James Caverlee}, {and} \bibinfo{person}{Fei Wang}.} \bibinfo{year}{2023}\natexlab{}.
\newblock \showarticletitle{Incorporating Time in Sequential Recommendation Models}. In \bibinfo{booktitle}{\emph{Proceedings of the 17th ACM Conference on Recommender Systems}}. \bibinfo{pages}{784--790}.
\newblock


\bibitem[Sun et~al\mbox{.}(2019)]%
        {bert4rec}
\bibfield{author}{\bibinfo{person}{Fei Sun}, \bibinfo{person}{Jun Liu}, \bibinfo{person}{Jian Wu}, \bibinfo{person}{Changhua Pei}, \bibinfo{person}{Xiao Lin}, \bibinfo{person}{Wenwu Ou}, {and} \bibinfo{person}{Peng Jiang}.} \bibinfo{year}{2019}\natexlab{}.
\newblock \showarticletitle{BERT4Rec: Sequential recommendation with bidirectional encoder representations from transformer}. In \bibinfo{booktitle}{\emph{Proceedings of the 28th ACM international conference on information and knowledge management}}. \bibinfo{pages}{1441--1450}.
\newblock


\bibitem[Taha and Malebary(2020)]%
        {creditcardfraud}
\bibfield{author}{\bibinfo{person}{Altyeb~Altaher Taha} {and} \bibinfo{person}{Sharaf~Jameel Malebary}.} \bibinfo{year}{2020}\natexlab{}.
\newblock \showarticletitle{An intelligent approach to credit card fraud detection using an optimized light gradient boosting machine}.
\newblock \bibinfo{journal}{\emph{IEEE access}}  \bibinfo{volume}{8} (\bibinfo{year}{2020}), \bibinfo{pages}{25579--25587}.
\newblock


\bibitem[Vaswani(2017)]%
        {vaswani}
\bibfield{author}{\bibinfo{person}{A Vaswani}.} \bibinfo{year}{2017}\natexlab{}.
\newblock \showarticletitle{Attention is all you need}.
\newblock \bibinfo{journal}{\emph{Advances in Neural Information Processing Systems}} (\bibinfo{year}{2017}).
\newblock


\bibitem[Xia et~al\mbox{.}(2023)]%
        {transact}
\bibfield{author}{\bibinfo{person}{Xue Xia}, \bibinfo{person}{Pong Eksombatchai}, \bibinfo{person}{Nikil Pancha}, \bibinfo{person}{Dhruvil~Deven Badani}, \bibinfo{person}{Po-Wei Wang}, \bibinfo{person}{Neng Gu}, \bibinfo{person}{Saurabh~Vishwas Joshi}, \bibinfo{person}{Nazanin Farahpour}, \bibinfo{person}{Zhiyuan Zhang}, {and} \bibinfo{person}{Andrew Zhai}.} \bibinfo{year}{2023}\natexlab{}.
\newblock \showarticletitle{Transact: Transformer-based realtime user action model for recommendation at pinterest}. In \bibinfo{booktitle}{\emph{Proceedings of the 29th ACM SIGKDD Conference on Knowledge Discovery and Data Mining}}. \bibinfo{pages}{5249--5259}.
\newblock


\bibitem[Yin et~al\mbox{.}(2020)]%
        {tabert}
\bibfield{author}{\bibinfo{person}{Pengcheng Yin}, \bibinfo{person}{Graham Neubig}, \bibinfo{person}{Wen-tau Yih}, {and} \bibinfo{person}{Sebastian Riedel}.} \bibinfo{year}{2020}\natexlab{}.
\newblock \showarticletitle{TaBERT: Pretraining for joint understanding of textual and tabular data}.
\newblock \bibinfo{journal}{\emph{arXiv preprint arXiv:2005.08314}} (\bibinfo{year}{2020}).
\newblock


\bibitem[Yuan et~al\mbox{.}(2020)]%
        {cntxt-awr}
\bibfield{author}{\bibinfo{person}{Weihua Yuan}, \bibinfo{person}{Hong Wang}, \bibinfo{person}{Xiaomei Yu}, \bibinfo{person}{Nan Liu}, {and} \bibinfo{person}{Zhenghao Li}.} \bibinfo{year}{2020}\natexlab{}.
\newblock \showarticletitle{Attention-based context-aware sequential recommendation model}.
\newblock \bibinfo{journal}{\emph{Information Sciences}}  \bibinfo{volume}{510} (\bibinfo{year}{2020}), \bibinfo{pages}{122--134}.
\newblock


\bibitem[Zha et~al\mbox{.}(2023)]%
        {tablgpt}
\bibfield{author}{\bibinfo{person}{Liangyu Zha}, \bibinfo{person}{Junlin Zhou}, \bibinfo{person}{Liyao Li}, \bibinfo{person}{Rui Wang}, \bibinfo{person}{Qingyi Huang}, \bibinfo{person}{Saisai Yang}, \bibinfo{person}{Jing Yuan}, \bibinfo{person}{Changbao Su}, \bibinfo{person}{Xiang Li}, \bibinfo{person}{Aofeng Su}, {et~al\mbox{.}}} \bibinfo{year}{2023}\natexlab{}.
\newblock \showarticletitle{Tablegpt: Towards unifying tables, nature language and commands into one gpt}.
\newblock \bibinfo{journal}{\emph{arXiv preprint arXiv:2307.08674}} (\bibinfo{year}{2023}).
\newblock


\bibitem[Zhang et~al\mbox{.}(2023)]%
        {fata-trans}
\bibfield{author}{\bibinfo{person}{Dongyu Zhang}, \bibinfo{person}{Liang Wang}, \bibinfo{person}{Xin Dai}, \bibinfo{person}{Shubham Jain}, \bibinfo{person}{Junpeng Wang}, \bibinfo{person}{Yujie Fan}, \bibinfo{person}{Chin-Chia~Michael Yeh}, \bibinfo{person}{Yan Zheng}, \bibinfo{person}{Zhongfang Zhuang}, {and} \bibinfo{person}{Wei Zhang}.} \bibinfo{year}{2023}\natexlab{}.
\newblock \showarticletitle{Fata-trans: Field and time-aware transformer for sequential tabular data}. In \bibinfo{booktitle}{\emph{Proceedings of the 32nd ACM International Conference on Information and Knowledge Management}}. \bibinfo{pages}{3247--3256}.
\newblock


\bibitem[Zhao et~al\mbox{.}(2023)]%
        {tabula}
\bibfield{author}{\bibinfo{person}{Zilong Zhao}, \bibinfo{person}{Robert Birke}, {and} \bibinfo{person}{Lydia Chen}.} \bibinfo{year}{2023}\natexlab{}.
\newblock \showarticletitle{Tabula: Harnessing language models for tabular data synthesis}.
\newblock \bibinfo{journal}{\emph{arXiv preprint arXiv:2310.12746}} (\bibinfo{year}{2023}).
\newblock


\end{thebibliography}


\appendix
\end{document}